\documentclass[12pt]{article}
\usepackage[utf8]{inputenc}
\usepackage{fancyvrb}
\usepackage{amsmath}
\usepackage{latexsym}
\usepackage{amsfonts}
\usepackage{centernot}
\usepackage[normalem]{ulem}
\usepackage{array}
\usepackage{amssymb}
\usepackage{graphicx}
\usepackage[backend=biber,
style=numeric,
sorting=none,
isbn=false,
doi=false,
url=false,
]{biblatex}

\usepackage{newunicodechar}
\newunicodechar{ﬁ}{fi}
\newunicodechar{ﬀ}{ff}

\usepackage{subfig}
\usepackage{wrapfig}
\usepackage{wasysym}
\usepackage{enumitem}
\usepackage{adjustbox}
\usepackage{ragged2e}
\usepackage[svgnames,table]{xcolor}
\usepackage{tikz}
\usepackage{longtable}
\usepackage{changepage}
\usepackage{setspace}
\usepackage{hhline}
\usepackage{multicol}
\usepackage{tabto}
\usepackage{float}
\usepackage{multirow}
\usepackage{makecell}
\usepackage{fancyhdr}
\usepackage[toc,page]{appendix}
\usepackage[hidelinks]{hyperref}
\usetikzlibrary{shapes.symbols,shapes.geometric,shadows,arrows.meta}
\tikzset{>={Latex[width=1.5mm,length=2mm]}}
\usepackage{flowchart}\usepackage[paperheight=11.0in,paperwidth=8.5in,left=1.0in,right=1.0in,top=1.0in,bottom=1.0in,headheight=1in]{geometry}
\usepackage[utf8]{inputenc}
\usepackage[T1]{fontenc}
\TabPositions{0.5in,1.0in,1.5in,2.0in,2.5in,3.0in,3.5in,4.0in,4.5in,5.0in,5.5in,6.0in,}

\setlistdepth{9}
\renewlist{enumerate}{enumerate}{9}
		\setlist[enumerate,1]{label=\arabic*)}
		\setlist[enumerate,2]{label=\alph*)}
		\setlist[enumerate,3]{label=(\roman*)}
		\setlist[enumerate,4]{label=(\arabic*)}
		\setlist[enumerate,5]{label=(\Alph*)}
		\setlist[enumerate,6]{label=(\Roman*)}
		\setlist[enumerate,7]{label=\arabic*}
		\setlist[enumerate,8]{label=\alph*}
		\setlist[enumerate,9]{label=\roman*}

\renewlist{itemize}{itemize}{9}
		\setlist[itemize]{label=$\cdot$}
		\setlist[itemize,1]{label=\textbullet}
		\setlist[itemize,2]{label=$\circ$}
		\setlist[itemize,3]{label=$\ast$}
		\setlist[itemize,4]{label=$\dagger$}
		\setlist[itemize,5]{label=$\triangleright$}
		\setlist[itemize,6]{label=$\bigstar$}
		\setlist[itemize,7]{label=$\blacklozenge$}
		\setlist[itemize,8]{label=$\prime$}

\begin{document}
\setlength{\parskip}{0.0pt}
\begin{adjustwidth}{0.5in}{0.0in}
\begin{justify}
{\fontsize{13pt}{15.6pt}\selectfont \textbf{The $``$cardinality of extended solution set$"$  criterion for establishing the intractability of \textit{NP} problems}\par}
\end{justify}\par

\end{adjustwidth}

\begin{adjustwidth}{0.5in}{0.0in}
\begin{justify}
{\fontsize{13pt}{15.6pt}\selectfont \textbf{ }\par}
\end{justify}\par

\end{adjustwidth}

 \begin{center}Arun U \end{center}
 \begin{center}arunu@alum.iisc.ac.in \end{center}
 
\begin{adjustwidth}{0.5in}{0.0in}
\begin{justify}
{\fontsize{13pt}{15.6pt}\selectfont \textbf{ }\par}
\end{justify}\par

\end{adjustwidth}

\begin{adjustwidth}{0.5in}{0.0in}
\begin{justify}
\textbf{Abstract: }The intractability of any problem and the randomness of its solutions have an obvious intuitive connection. However, the challenge till now has been that there is no practical way to firmly establish if the solution to a problem is actually random (or whether it has some hidden undiscovered structure, which upon being detected would render it non-random). This has prevented the conclusive declaration of hard problems (such as \textit{NP}) as being definitely intractable. For dealing with this, a concept called "extensibility" of a sequence is developed. Based on this, a criterion termed as $``$cardinality of extended solution set$"$ is conceived to ascertain the (non)randomness of any sequence. Further, this can then be used to establish the (in)tractability of any problem depending on whether its solutions are random or non-random. This criterion is applied to problems such as \textit{2-SAT}, \textit{3-SAT} and hardness of approximation to analyze their (in)tractability. Finally, a proof for the validity of the Unique Games Conjecture based on the same criterion is also presented.
\end{justify}\par

\end{adjustwidth}

\begin{adjustwidth}{0.5in}{0.0in}
\begin{justify}
\end{justify}\par

\end{adjustwidth}

\vspace{\baselineskip}
\begin{enumerate}
	\item \textbf{Introduction}\par

Intuition will lead one to believe that there must be a connection between computability, tractability and randomness. But, given that randomness has been a field of study across various branches such as mathematics, statistics and theoretical computer science, it begs the question as to why there are still no definitive means of classifying problems as being tractable or intractable. The computability of real numbers was indeed a prime motivation for Turing himself as captured in his original paper (Turing [1937]) and computable analysis as a field has since developed into an active area of research (e.g. - Di Gianantonio[1993], Weihrauch [1995], Avigad[2014]). However, the question of whether there is a practical means of ascertaining if a given problem maps to the space of computable or uncomputable real numbers is itself an unanswered one till now. To resolve this will be the key objective of this work, which will be largely based on developing a deeper understanding of randomness.\par

Traditional randomness tests have hitherto been based on solely analysing statistical parameters associated with a particular sequence. There are indeed several such tests e.g. Wald-Wolfowitz runs test (Wald [1940]), Kolmogorov-Smirnov test (Massey [1951]), Maurer's universal statistical test (Maurer [1992]), Coron's test (Coron [1998]) etc. But, if these had been conclusive in establishing the randomness of a sequence, then it should have also been possible to identify problems that are intractable with relative ease since any problem with truly random solutions is almost certainly going to be intractable. The fact that one is unable to do so is indicative of the difficulty in coming up with a fool proof test of randomness.\par

Further, the connection between intractability and uncomputability also appears intuitively obvious. Any attempt to definitively establish the intractability of a problem would most likely involve showing that its solution would lie in the uncomputable real numbers space. However, it has been shown that there are examples of so called $``$absolutely normal numbers$"$, which can be deemed to be completely $``$random$"$  by any statistical measure, which are also computable (Becher and Figueira [2002]). This again underscores the inadequacy of purely statistical measures for measuring randomness - something that will be delved into more deeply in this paper. This is also buttressed by the fact that pseudo-random generating algorithms can generate random sequences, which will qualify as being random by all such statistical tests, but by their very nature are computable (since they are run on computers!). All these point to a gap in the current understanding of randomness. In particular, one of the central arguments of this paper is that there is a need to distinguish between statistical randomness and computational randomness. While the former can be analyzed through "fuzzy" parameters (e.g. the distribution of digits), the latter has to be definitive.\par

The approach adopted herein to build the concept of computational randomness and employ the same to prove the intractability of \textit{NP} problems is as follows:
\begin{enumerate}
	\item First, it is proven that the cardinality of the set of all computationally non-random sequences is countably infinite, while the cardinality of the set of all computationally random sequences is uncountably infinite\par

	\item It is then demonstrated that for a finite sequence to be non-random, it also necessarily has to be "extensible" (based on a prior definition of extensible which is undertaken) \par

	\item It is then shown that the cardinality of the "extended set" of a sequence can be used to ascertain its (in)extensibility (and (non)randomness) \par

	\item It is finally concluded that if the extended set of a sequence is uncountable, the sequence itself is random and the problem for which it is a solution is intractable\end{enumerate}\par

This is then extended to understand the hardness of approximation problems and finally arriving at a proof of the Unique Games Conjecture (UGC) as follows:
\begin{enumerate}

    \item 
    It is first demonstrated that for a particular reduced level of accuracy, the recast decision problem (of an originally intractable exact problem) is equivalent to the trivial case of a problem whose solutions include both random and non-random sequences. \par
    
    \item
    Hence, to analyze the (in)tractability of approximate problems, one needs to focus on the optimization case (rather than the decision case). As such,  the same principles of extensibility can be utilized, but applied on sequences of variable evaluations for the optimization problem (rather than variable assignments as with the decision problem).\par
    
    \item
    This principle is then applied to the case of \textit{MAX-3SAT} and unique label cover problem. In so doing, it is proved that the former is approximable while the latter is not, thus proving the UGC. \par
    
    \end{enumerate}\par

\vspace{\baselineskip}
	\item \textbf{Definitions}\par
\textbf{\textit{Definition 2.1 (Random sequence):}} A sequence  \( S \)  is said to be a random sequence if for any pair of elements, \textit{s\textsubscript{i}} \textit{s\textsubscript{j}} $\in$ \(S\), there is no mapping from $s_i$ to $s_j$. In other words, $\nexists$ any function \(f\) : $\mathbb{N} \times \mathbb{N} \times \mathbb{N}\rightarrow \mathbb{N} $ such that $f(s_i,i,j)=s_j$ (where $s_i$ and $s_j$ are the \textit{i}\textsuperscript{th} and \textit{j}\textsuperscript{th} elements of \(S\)).  \\Hence, any sequence which is not a random sequence is a non-random sequence. In such a sequence, for all pairs of elements $\textit{s\textsubscript{i}}, \textit{s\textsubscript{j}} \in$ \(S\), $\exists$ a mapping between  \textit{s\textsubscript{i}} and \textit{s\textsubscript{j}}.\par

\begin{justify}
\textbf{\textit{Definition 2.2 (k-sequence):}} A sequence  \( S \)  is said to be a \textit{k}-sequence if each element
of \textit{S} can take one of \textit{k} values (where $\textit{k} \in \mathbb{N})$. In particular, a \textit{2}-sequence is one where every element of the sequence can have only one of two possible values (e.g. 0 or 1).\end{justify}\par

\begin{justify}
\textbf{\textit{Definition 2.3 (Computable sequence):}}   A sequence  \( S \) of length \(l\) is said to be computable if $\exists$ a function \textit{f}$\colon\ \mathbb{N}\rightarrow  \mathbb{N}$ such that for 0\textless\(n\)$\leq$ \(l\), \textit{f}(\(n\))= \(s\)\textsubscript{\(n\)},  where \(s\)\textsubscript{\(n\)} is the \(n\)\textsuperscript{th} element of \(S\).\end{justify}\par

\vspace{\baselineskip}

\item \textbf{Non-random sequence generation}\par

Let the following thought experiment be undertaken. Consider a computer program, which generates a non-random \(k\)-sequence of any required length  in less than exponential time.  Let such a program be designated as a $``$non-random sequence generator$"$. Let $ \Gamma $  be the set of all such possible non-random sequence generators. E.g., if one were to consider only \textit{2}-sequences, $ \Gamma $  will have all possible programs which will can generate every possible non-random \textit{2}-sequence conceivable.\   

Now, the non-random sequence generators could be of different types w.r.t. their time complexity. E.g. a sequence generator could be of \(O(n)\) or \(O(log~n)\) etc. So, let $ \Gamma_1 $, $ \Gamma_2 $ etc. be the sets of different non-random sequence generators where the programs in each of these sets belong to the same time complexity. E.g. $ \Gamma_1 $ could be the set of all programs of complexity \(n\), $ \Gamma_2 $ could be the set of all programs of complexity \(log~n\) and so on.\par

Hence, it is clear that $\Gamma = \underset{i=1}{\overset{k }{\bigcup \Gamma_i}}$, where \(k\) is the number of all such sets of non-random sequence generators.

\vspace{\baselineskip}

\textbf{\textit{Theorem 3.1. The cardinality of the set of all non-random k-sequences is countably infinite.}}\par

\textbf{\textit{Proof.} }For any set of non-random sequence generators $ \Gamma_i $,  some logic can be devised by which the sequence generators can be sorted in a particular order. E.g. one could order it by the length of the programs themselves. For this, assume that the programs would be written in some programming language. So, the programs could be ordered by the number of characters in them. In case of two programs having the same length, they could be further ordered lexicographically on the basis of the first character, second character etc. (i.e. similar to words arranged in a dictionary). In this way one could convert $ \Gamma_i $  to an ordered set $ \Gamma_i  ^{'}$ consisting of programs \textit{P\textsubscript{1}}, \textit{P\textsubscript{2}}, \textit{P\textsubscript{3}}$ \ldots $ (A similar technique has been used by [Gusfield] for coming up with a simple proof for a variant of the G{\"o}del's theorem).\par 
$\therefore$ the cardinality of $ \Gamma_i $  is countably infinite i.e. $\aleph$\textsubscript{0} (as its members have been mapped to the natural numbers). \par
Now, the sequences of different lengths that can be generated by each of the programs in $\Gamma_i^{'}$ can themselves be arranged in increasing order of their lengths.\par
Since $\mathbb{N} \times \mathbb{N}$ is countable (Tao[2009]), \par
$\Rightarrow$ set of all sequences that can be generated by $\Gamma_i$ is countable.\par
Now, since the union of countable sets is also countable,\par
$\Rightarrow$ set of all non-random \(k\)-sequences (generated by all programs in $\Gamma$) is countable. \par
Since this is an important result, it is again reinforced later through computability arguments (ref. Theorem 6.1, Theorem 6.2,  Theorem 6.3 and  Theorem 6.4). 

\vspace{\baselineskip}
\textbf{\textit{Theorem 3.2. The cardinality of the set of all random k-sequences is uncountably infinite.}} \par

\textbf{\textit{Proof.}} Now, since the sequences are random, for any pair of elements, say $s_i$ and $s_j$, there is no unique mapping between them. Hence, this implies that for any value of $s_i$, $s_j$ has to take more than one value (else $s_i$ will be uniquely mapped to $s_j$ and hence the sequence will be non-random). Hence, for any required length \(n\), the cardinality of the set of all  random sequences  will be at least $2^n$. Further, as the lengths of the sequences become arbitrarily large, the set of all the sequences becomes uncountable (Tao[2009]) (i.e.  its cardinality is $\aleph$\textsubscript{1}).\par

\vspace{\baselineskip}
Here, it is also worth clarifying as to why the set of all non-random generators, $\Gamma$ excludes the programs that run in exponential time. For this, consider the sequence generator, which involves arbitrary assignment of values to each element of the sequence.  Such a guessing program can generate any particular \(k\)-sequence in exponential time, but it does so not by leveraging the relationships between the elements of the sequence (which is the requisite for a non-random sequence generator) but by random assignment.  Hence, such programs are excluded from $\Gamma$ as this will have implications on deciding the tractability of any problem as will be demonstrated later.  \par

\vspace{\baselineskip}
\textbf{\textit{Theorem 3.3. If a set contains sequences which are all either random or all are non-random, and if the set is uncountable, then all the sequences have to be random}}. \par
\textbf{\textit{Proof.}} Here the key condition that is imposed is that all the sequences in the set are either random or all are non-random (i.e. there cannot be a mix of both random and non-random sequences).
So, consider the case where they are all non-random. Then, the maximum cardinality which is achievable from Theorem 3.1 is $\aleph$\textsubscript{0}, (which is less than $\aleph$\textsubscript{1}). Hence, all the sequences have to necessarily be random.

\vspace{\baselineskip}
	\item \textbf{Extensibility}\par

Thus far, sequences which belong to particular sets (i.e. random or non-random) have been analysed. But, consider the case of individual sequences of finite length (whose set related details are not known). E.g. consider the following sequence of length 10:\par

1874903624\  \par

Although, the above sequence "appears" to be random, it is hard to say definitively whether it is actually random or non-random. However, the sequence below can be definitely said to be non-random if the pattern of the above 10 elements is repeated forever:\par

1874903624\uline{\textcolor[HTML]{FF0000}{1874903624}}1874903624\par

On the other hand, the following sequence may appear to be non-random (as it alternates between 0 and 1):\par

0101010101\par

But, on extending it, if it looks like this:\par

0101010101\uline{\textcolor[HTML]{FF0000}{110100101100}}\par

then, it may no longer be very obvious if the above sequence is indeed non-random. What this demonstrates is that unless one has information about how the sequence behaves when extended indefinitely, it cannot be said with certainty whether it is indeed random or non-random. Further, even one non-random sequence cannot be distinguished from another when they both are exactly the same for a finite length. E.g., consider the following sequence which is supposed to be non-random:\par

12345678\par

The above sequence could be a part of several non-random sequences, e.g.:\par

12345678\uline{\textcolor[HTML]{FF0000}{12345678}}12345678\par

123456789\uline{\textcolor[HTML]{FF0000}{123456789}}123456789\par

1234567890\uline{\textcolor[HTML]{FF0000}{1234567890}}1234567890\par

etc.\par

This implies that in order to ascertain the non-randomness of a finite sequence, it is not enough to map it to a computer program which can generate it. This is because the finite sequence could also completely match the initial part of a random sequence (which would be of infinite length). In order to address this, the concept of extensibility is developed hereunder. \par

\begin{justify}
\textbf{\textit{Definition 4.1 (Extensible):}} 
Consider a sequence  \( S_{l} \) of length  \( l  \)  which is a solution to an instance of a problem $ \theta $.  Now, \( S_{l} \) can be characterized as being "extensible" if  $\exists$ a program \textit{P\textsubscript{i}}, which runs in less than exponential time and fulfils the following two criteria:
\\
\textbf{\textit{Condition i)}}  \textit{P\textsubscript{i}} can generate  \( S_{l} \) 
\\
\textbf{\textit{Condition ii)}} Any other string  \( S_{m} \)  of length \textit{m} $\in \mathbb{N}$, which is generated by \textit{P\textsubscript{i}} is also a solution of an instance of $ \theta$
\\
\\ If the above two conditions are met, one can be sure that \( S_{l} \) is mapped to \textit{P\textsubscript{i}} in the context of the problem for which it is a solution and hence non-random. It also follows that \textit{P\textsubscript{i}} will be a non-random sequence generator (and $\in \Gamma$).
\end{justify}

\begin{justify}
\textbf{\textit{Definition 4.2 (Extended sequence):}} 
 An "extended sequence" \( S_{m} \) of a sequence \( S_{l} \), which is a solution to an instance of a problem $\theta$ is one that fulfills the following criteria:
 \\
\textbf{\textit{Condition i)}} The length of \( S_{m} \) (i.e. \(m\)) is greater than that of \( S_{l} \) (i.e. \(l\)) 
\\
\textbf{\textit{Condition ii)}} The first \(l\) elements of \( S_{m} \) are the same as the corresponding elements of \( S_{l} \)  
\\
\textbf{\textit{Condition iii)}} \( S_{m} \) is also a solution to an instance of $\theta$
\end{justify}

\begin{justify}
\textbf{\textit{Definition 4.3 (Extended set):}} 
The set of all extended sequences of a particular sequence is called its "extended set".
\end{justify}

The rationale for adopting these definitions is that for reasons discussed above, analyzing individual solution sequences of any particular problem is of little help when it comes to establishing its (in)tractability (as it is impossible to determine their (non)randomness in isolation). Hence, what is needed is to understand the inherent nature of the entire solution space (i.e. its (non)randomness), which in turn is also inherited by the individual sequences. The concept of extensibility as defined above helps achieve that because if there is a generator algorithm, which can be identified such that the sequences it generates are solutions to different instances of the problem, then it definitively captures the structural characteristic of the problem's solution space.  
\\The question that can now be asked is - \textit{$``$What is the exact connection between extensibility and randomness?$"$}. This is answered in the next Section. The particular area of interest is in Constraint Satisfaction Problems with growing number of variables.

\vspace{\baselineskip}
	\item \textbf{Relation between the extended set of a solution sequence of a problem and tractability of the problem}

From the previous discussion, it is obvious that it is not possible to ascertain the (non)randomness of any finite sequence by just analysing it in isolation. Hence, in such a case, the way one could ascertain the (non)randomness of the sequence is to analyse its extended set using the following results. \par
\vspace{\baselineskip}
\textbf{\textit{Theorem 5.1. A problem whose solution takes the form of a k-sequence, the  extended set of which grows at rate of k\textsuperscript{n} members will have random solutions}}.
\\
\textbf{\textit{Proof.}} Consider a problem $\theta$, an instance of which i.e. $\theta$\textit{\textsuperscript{j}} has a solution \( S_{l} \), which is a \textit{k}-sequence of length \textit{l}. Now, let the cardinality of the extended set of \( S_{l} \) grow at the rate of \textit{k\textsuperscript{n}}. The task here is to ascertain if \( S_{l} \) is random or non-random.
\\ Now, as the length of the extended sequences of \( S_{l} \) increase to cover all natural numbers $\mathbb{N}$, its extended set becomes uncountable (Tao[2009]).\par 
Further, by using the principle of symmetry all the sequences in the extended set will either be random or non-random. This is because the growth in the solution space is engendered by an increasing number of variables subject to similar constraints. Hence, it should have similar outcomes in terms of either random or non-random sequences. This is because owing to the principle of symmetry, it cannot be the case that certain constraints produce random sequences, while others produce non-random sequences. But, from Theorem 3.3, since the extended set itself is uncountable, the sequences have to be random. Hence, the solutions of $\theta$ are random.
\par

\vspace{\baselineskip}
\textbf{\textit{Theorem 5.2:}} \textbf{\textit{A problem whose solution takes the form of a k-sequence, the extended set of which grows at rate of k\textsuperscript{n }is intractable}}.
\\
\textbf{\textit{Proof. }}Assume that the problem (designated as $\theta$) is tractable.
\par

 \(  \therefore   \exists  \) a program \( P \) that can compute the solution sequence  \( S \)  for any instance of $\theta$ and also the sequences in the extended set of \( S \) in less than exponential time.
\\$\Rightarrow$ \( S \) is non-random 
\\But, this contradicts Theorem 5.1. 
\\Hence, no such program exists and $\theta$ is not tractable (i.e. it cannot be solved in less than exponential time). \par
This result is again reinforced later through computability arguments in the next Section (ref. Theorem 6.1, Theorem 6.2, Theorem 6.3 and Theorem 6.4).   

\vspace{\baselineskip}

	\item \textbf{Computability and extensibility}
\par
The focus of this Section is on computablity and its relationship with randomness and extensibility. 
\vspace{\baselineskip}

\textbf{\textit{Theorem 6.1: A random k-sequence is uncomputable.}}
\\
\textbf{\textit{Proof.}} Let the random \(k\)-sequence be designated as \( S \). Since  \( S \)  is random, $\nexists$ any function \(f\) : $\mathbb{N} \times \mathbb{N} \times \mathbb{N}\rightarrow \mathbb{N} $ which takes as inputs, the index and value of an element in \( S \) (say \textit{s\textsubscript{i}}) and also the index of another element (say \textit{s\textsubscript{j}}) and maps them to the latter's value.
\\
Now, let  \( S \)  be computable.
\\
\(  \Rightarrow   \exists  \)  function \(p\) $\ni$  $p\left(i\right)=m$ and   $p\left(j\right)=n$ (where \(m\) and \(n\) are the values of the \(i\)\textsuperscript{th} and \(j\)\textsuperscript{th} elements of \(S\) respectively).
\\With this, one can compose a new function \(q\) which takes as its inputs, (\textit{i},$p\left(i\right)$,\textit{j}) and maps them to $p\left(j\right)$.  
\\
$\Rightarrow$  \( S \)  is not random, which is a contradiction.
\\
Hence,  \( S \)  is uncomputable.\par

\vspace{\baselineskip}

\textbf{\textit{Theorem 6.2: An uncomputable k-sequence is 
random.}}\par
\textbf{\textit{Proof.}} Let the uncomputable \(k\)-sequence be designated as \( S \). \par
Now, if \( S \) is non-random, $\Rightarrow \exists$ function \(f\) such that $f(i,s_i,j)=s_j$ where \textit{s\textsubscript{i}} and \textit{s\textsubscript{j}} are the \textit{i}\textsuperscript{th} and \textit{j}\textsuperscript{th} elements of \(S\) respectively.\par
$\Rightarrow \exists$ function \(g\) such that $g(i)=s_i$ (for else, it would not have been possible for \(f\) to have \(i\) and $s_i$ as inputs). \par
$\Rightarrow$ \(S\) is computable, which is a contradiction. \par
Hence, \(S\) has to be random.

\vspace{\baselineskip}

\textbf{\textit{Theorem 6.3: The cardinality of set of all computable sequences is countably infinite }}\par
\textbf{\textit{Proof.}} For proving this, a sorting algorithm which sorts the set of computable sequences is constructed as follows.\par
Start with any of the sequences which becomes the first of the sorted list and is designated as \(S\)\textsubscript{1}. Then, the sorting of remaining sequences is done based on the maximum number of elements at the beginning of each sequence that match with the corresponding elements of   \(S\)\textsubscript{1}. E.g. assume that the first element of the first sequence that is compared with first element of \(S\)\textsubscript{1} does not match. Then, the maximum matching count for that sequence is 0 and maximum matching count for all sequences compared till that point is also 0 (since no other sequence has been compared with \(S\)\textsubscript{1} so far). Hence, this sequence becomes the second one on the sorted list.\par
But, now say the next sequence which is compared with \(S\)\textsubscript{1} has the first 2 elements in common. Hence, this will now become the second in the sorted list and the previous sequence gets pushed down. \par
This exercise can be repeated for all sequences. In case two sequences have the same number of starting elements matching, the next set of matching elements (i.e. after the interim non-matching elements) can be compared and again sorted on that basis. This procedure can be done iteratively if the second set of matching elements also turn out to be the same etc.\par
This procedure will produce a sorted list of the sequences which implies that the set of all computable sequences is countable. Note that this approach can be undertaken for computable sequences as there is a defined mapping between the index and value of each element of the sequence.   \par

\vspace{\baselineskip}
\textbf{\textit{Theorem 6.4. The cardinality of the set of all uncomputable sequences is uncountably infinite.}} \par

\textbf{\textit{Proof.}} This can be proven through the familiar Cantor diagnolization technique applied on the set of computable sequences. Since the set of all the computable sequences is countable, they can be ordered (using a methodology as described in Theorem 6.3) as say, \(S\)\textsubscript{1}, \(S\)\textsubscript{2}, \(S\)\textsubscript{3} etc. Now, by reversing the successive elements of the sequences \(S\)\textsubscript{1}, \(S\)\textsubscript{2}, \(S\)\textsubscript{3} etc. starting from the first element of \(S\)\textsubscript{1}, second element of \(S\)\textsubscript{2}, third element of \(S\)\textsubscript{3} etc., a new sequence  which does not belong to any of these sequences can be produced. This implies that the set of all uncomputable sequences cannot be mapped to the set of natural numbers $\mathbb{N}$ and   hence, it should be uncountably infinite (i.e.  its cardinality is $\aleph$\textsubscript{1}).\par

\vspace{\baselineskip}
Theorems 6.1, 6.2 and 6.3 together serve to validate Theorem 3.1. They along with Theorem 6.4 also validate Theorem 5.2 (since all random sequences are uncomputable and hence and by definition cannot be generated through any deterministic program which does not involve guessing the values of the elements). \par

What the above theorems also show is that computability and randomness go hand in hand. The relationship between randomness and extensibility has already been demonstrated earlier. Now, the same is done w.r.t. computability. One of the main aspects of computability as laid out by Turing is that it can only be applied to sequences which are of finite length. Hence, this poses a challenge when it has to be applied to a sequence, whose true nature is only revealed when it is extended to infinity (as is the case with random sequences). This is where the concept of extensibility becomes a handy tool for doing so. It helps analyze the behaviour of finite sequences (which lie within the framework of a Turing machine) by relating them to their nature when they are extended infinitely. Hence, extensibility helps plug an important gap in the traditional concept of computability.

\vspace{\baselineskip}
	\item \textbf{Application of cardinality of extensible set criterion to the Boolean Satisfiability problems}\par
In this Section, the criterion for ascertaining the (in)tractability of any problem is tested by applying it to the Boolean Satisfiability problems.
One of the well known facts about the Boolean Satisfiability problems is that while the \textit{2-SAT} problem is tractable, the \textit{3-SAT} problem is intractable. Hence, it is verified if the above criterion can explain the same.\par

\vspace{\baselineskip}

\textbf{\textit{Theorem 7.1. The 3-SAT problem is intractable.}}\par
\textbf{\textit{Proof.}}
Consider a  \textit{3-CNF}  formula  \( F  \) of size  \( n \)  of the form: 
\\ 
\(  \left( l_{1} \bigvee l_{2} \bigvee l_{3} \right)  \bigwedge \left( l_{4} \bigvee l_{5} \bigvee l_{6} \right)  \ldots   \left( l_{n-2} \bigvee l_{n-1} \bigvee l_{n} \right)  \).
\par
The logical implications of the clauses in this case will take the form of  \(  \sim l_{1} \Rightarrow l_{2} \bigvee l_{3} \)  etc.\  Hence,  \( F \)  can be equivalently represented as an implication graph of all these possibilities and solving  \( F \)  will entail traversing through such a graph.
\par

Now, for any of the clauses (say the first one), suppose for any particular instance of the problem, one is able to disregard one of the logical implications, i.e. one can assume that  \(  \sim l_{1} \Rightarrow l_{2} \)  (instead of  \(  \sim l_{1} \Rightarrow l_{2} \bigvee l_{3} \) ) and still able to ascertain the satisfiability of  \( F \). This can be so only if  \( l_{2}=1 \). Hence, the value of  \( l_{3} \)  will have no bearing in this case.
\par

Now, in another instance of the problem, let there be an additional constraint in the form of the following clause:  \(  \left(  \sim l_{1} \bigvee \sim l_{2} \bigvee l_{3} \right)  \). Now,  \( F \)  cannot be ascertained to be satisfiable till the value of  \( l_{3} \)  is also checked. This proves that across all instances,  all possible permutations of solutions can be covered by incorporating different varieties of constraints (in the form of the disjunctive clauses) in the input. This implies that the extended set grows at the rate of $2^n$ and hence from Theorem 5.2, the problem is intractable (i.e. cannot be solved in polynomial time).\par

\vspace{\baselineskip}
 \textbf{\textit{Theorem 7.2. The 2-SAT problem is tractable.}}\par
\textbf{\textit{Proof.}}
In contrast to the above, in  \textit{2-SAT}, each individual clause of the form  \( l_{1} \bigvee l_{2} \)  can be expressed in the form of  \( a \Rightarrow b \).  This is because since every disjunctive clause in the  \( CNF \)  has to be true, it follows that  \(  \sim  l_{1} \Rightarrow  l_{2} \)  and  \(  \sim  \)   \( l_{2} \Rightarrow  l_{1} \). Hence, this is completely deterministic because if  \( a \)  is true so is  \( b \)  and if  \( b \)  is false, so is  \( a \). In other words, there is only one choice for each literal depending on the value of the other literal. So, to check if a given  \textit{2-CNF}  formula is satisfiable, one only needs to check if there are any conflicting conditions for each of the literals in all the clauses it appears. In case any such conflicts are detected for a particular literal, then the whole formula will be unsatisfiable. Therefore, the total number of logical implications for the  \textit{2-SAT}  problem is only  \( 2n \)  which implies that its solutions are non-random. Hence, the problem is tractable.\par

\vspace{\baselineskip}
	\item \textbf{\textbf{Hardness of approximation }}\par
One of the intriguing aspects of finding approximate solutions to hard problems is the existence of well defined thresholds of approximability (beyond which the problem becomes intractable). This can also be understood from the viewpoint of extensibility. For this, consider the trivial problem (designated as $\theta $\textsuperscript{TRIV}) for which any sequence is a valid solution for any instance of the problem. Similarly, when a hard problem is recast to allow for  approximate solutions by relaxing the constraints, at a particular level of (reduced) accuracy, the decision problem will be annulled. In that sense, it will be similar to $\theta $\textsuperscript{TRIV} as both do not have the need to make any decisions and hence any sequence will be vaild solution for all instances of the problem. This is illustrated with the example of the \textit{MAX-3SAT} problem.\par    
(H{\aa}stad[2001]) has shown that the approximability threshold for the \textit{MAX-3SAT} problem is 7/8, which also corresponds to the random assignment case. That is because since each \textit{3-CNF} clause has 3 disjunctive literals, there are 8 permutations of the literals. Hence, the chances of any arbitrary assignment being right is 7/8, which obviates the need for any decision making. While random assignment is one obvious approach, other approaches also exist (e.g. semi-definite programming) which accomplish the same end for other problems.\par
In other words, if the requisite accuracy for a hard problem is set at any level below that of the approximation threshold, no discrete choice needs to be incorporated in any approach to solve the problem and the problem is one of pure optimization.  However, for any threshold greater than the threshold, it will involve two steps:
\\
\textbf{\textit{Step 1}}: Finding the solution space itself 
\\ \textbf{\textit{Step 2}}: Optimizing within the solution space 
\\And, as shown earlier since \textbf{\textit{Step 1}} itself is intractable, the whole problem becomes intractable. \par
This implies that the approximability threshold for any approximation problem is that accuracy level beyond which the decision problem resurfaces, which renders it intractable. But, the question that can now be asked is - \textit{"What defines the criterion for approximability?"}. This is discussed in the context of a proof for the UGC. But, before doing that the standard definitions pertaining to constraint satisfaction problems are adopted as under.

\vspace{\baselineskip}

\textbf{\textit{Definition 8.1 (Constraint Satisfaction Problem):}} A Constraint Satisfaction Problem (CSP) is defined as a triple $\langle X,D,C\rangle$, where 
\\$X=\{X_1, ...,X_n\}$ is a set of variables,
\\$D=\{D_1, ...,D_n\}$ is a set of their respective domains of values, and
\\$C=\{C_1, ...,C_m\}$ is a set of constraints.
Each variable $X_i$ can take on the values in the nonempty domain $D_i$. Every constraint  $C_j \in C$ is in turn a pair $\langle t_j,R_j \rangle$, where $t_j \subset X$ is a subset of \(k\) variables and $R_j$ is a \(k\)-ary relation on the corresponding subset of domains $D_j$.\par
\vspace{\baselineskip}

\textbf{\textit{Definition 8.2 (Evaluation):}} An "evaluation" of the variables is a function from a subset of variables to a particular set of values in the corresponding subset of domains. \par
\vspace{\baselineskip}

\textbf{\textit{Definition 8.3 (Satisfy):}} An evaluation \(v\) satisfies a constraint $\langle t_j,R_j\rangle$ if the values assigned to $t_j$ satisfies the relation $R_j$.  
\vspace{\baselineskip}

Now, it is known that the approximation problems are one of optimization (rather than decision) with the objective of maximizing the number of satisfied constraints in an evaluation. Hence, to ascertain their (in)tractability,
the same principles of extensibility can be used but, w.r.t. variable evaluations (in
contrast to variable assignments). In other words, all the results that were used to
ascertain the (in)tractability of exact solutions can be used for the approximate case
as well. But, instead of applying them to sequences of variable assignments for the decision case, they
would be applied to sequences of variable evaluations for the optimization case. This is demonstrated through
application to \textit{MAX-3SAT} and UGC as follows.

\vspace{\baselineskip}

 \textbf{\textit{Theorem 8.1. The MAX-3SAT problem is tractable.}}\par
\textbf{\textit{Proof.}}
Consider an evaluation sequence for an instance of a \textit{3-SAT} problem. Now, the task at hand would be to analyze the extended set of this evaluation sequence. For this, consider a variable sequence $S$ of length \(n\). It is clear that every constraint in the \textit{3-SAT} case can constrain 3 variables. Hence, the total possible combinations of constraints that can constrain this variable sequence will be ${n \choose 3}$. Further, since each constraint constrains 3 variables, the total number or evaluations will be $3 \times {n \choose 3} $. From the criterion of extensibility, as the lengths of the sequences $\rightarrow \infty$, ${n \choose 3}=O(n^3)$. Hence, the cardinality of the extended set of evaluations for any evaluation sequence will be $O(3n^3)=O(n^3)=\aleph\textsubscript{0}$. Therefore, the problem is tractable.

\vspace{\baselineskip}
Now, the focus is shifted to the UGC. The UGC formulated by Khot[2002] hypothesizes that for arbitrarily small constants $\epsilon, \delta >0$, there exists a constant \(k\)=\(k\)($\epsilon, \delta$) such that it is \textit{NP}-hard to determine if a unique 2-prover game with answers from a domain of size \(k\) has value at least 1-$\epsilon$ or at most $\delta$. This is equivalently stated as a unique label cover problem where the variables are treated as the vertices of the graph and the constraints as its edges. An attempt is made herein to prove the UGC. 
\vspace{\baselineskip}

 \textbf{\textit{Theorem 8.2. The Unique Games Conjecture is valid.}}\par
\textbf{\textit{Proof.}} It is easy to see that the exact case of the unique label cover problem is an easy one to solve (i.e. it is in \(P\)). Since each vertex is uniquely mapped to \(k\) values, it is clear that for a sequence of variables of length \(n\), there are a maximum of  $kn$ computations to ascertain the satisfiability of the problem.  But, one of the unresolved questions about the unique label cover problem is the exact source of the hardness when it comes to the approximate case. E.g. it is sometimes attributed to the arbitrarily large domain sizes of the attributes which result in multiple combinations for the approximate case. But the paradox is that w.r.t. the variable assignments problem, if an algorithm is able to solve for the exact case, it appears counter intuitive to assume that it won't be able to solve the approximate case (as it is a relaxation of the accuracy requirement). This is also the view that emerges from the extensibility criterion. This is because if the cardinality of extended set of computations for the exact case is $kn$, then it is $kn - x$ (where \(x\) is some number which depends on the reduced level of accuracy). Hence, there doesn't appear to be any basis for this to be a cause of the hardness for the approximate case. \par  

\noindent Therefore, focus needs to be shifted to the variables evaluation space in the context of the optimization problem in order to ascertain the (in)tractability of the approximation problem. For this, the number of constraints applicable to any variable sequence needs to be computed just as in the \textit{MAX-3SAT} case. In the case of \textit{MAX-3SAT}, there is just one type of constraint, i.e.- the disjunction of 3 variables (or literals). However, the unique label cover problem is different in this respect. If there are \(k\) values that each variable can take, then there can be a total of $k^2$ possible constraints. E.g. let the domain of the variables be \{a,b,c\}. Then the constraints can be any member of \{aa,ab,ac...ca,cb,cc\}. Hence, there will be a total of $3^2=9$ constraints. Further, each constraint acts on 2 vertices (or 1 edge) of the unique label cover graph. Hence, the total combinations of constraints for a variable sequence of length \(n\) will be $k^{2^{n \choose 2}}$ and the total number of evaluations will be $1\times k^{2^{n \choose 2}}=k^{2^{n \choose 2}}$ (since each constraint acts on 1 edge).
\\
Now, as \(n\)$\rightarrow \infty$, ${n \choose 2}=O(n^2)$.
Hence, as \(n\)$\rightarrow \infty$, the cardinality of extended set of evaluations  for any evaluation sequence will be $\aleph\textsubscript{1}$. Hence, from Theorem 5.2, the unique label cover problem is inapproximable, or the UGC is valid. However, the important point to note is that the UGC would be valid even if the domain size is a fixed number (>1). Hence, there is no dependence of the UGC to the arbitrarily large domain sizes.

\vspace{\baselineskip}
	\item \textbf{Conclusion}
\par

There are five key takeaways from this paper: 

\begin{enumerate}[topsep=0pt]
\item There is a difference between statistical and computational randomness. While the former can be independently analyzed w.r.t. any sequence, the latter requires the entire context of the problem (an instance of which, the sequence is a solution).

\item As such, computability and computational randomness are not universal or absolute. Rather, they need to be understood in the context of the problem itself. The exact same (finite) sequence could be characterized as computable and non-random in the context of one problem, while it could be characterized as uncomputable and random in the context of another problem.

\item The true nature of a solution sequence (i.e. its (un)computability and its (non) randomness) is only revealed when it is extended to infinity. But, the traditional concept of Turing computability is equipped to deal only with finite sequence outputs. Hence, the concept of extensibility that has been introduced here becomes indispensible to bridge this gap in analyzing the characteristics of finite and infinite  sequence outputs.

\item Extensibility helps resolve the hitherto unsolved problem of establishing the intractability of \textit{NP} problems by demonstrating the random nature of their solutions. It does so on the basis of the randomness of the entire solution space, which in turn would also imply the randomness of the individual solutions.

\item The Unique Games Conjecture is valid (but there is no dependence of its validity on the arbitrarily large nature of the domain size of the variables).    

\end{enumerate} 

\end{enumerate}\par

\vspace{\baselineskip}
\newpage

\vspace{\baselineskip}\begin{adjustwidth}{0.4in}{0.0in}
\begin{justify}
\textbf{REFERENCES}
\end{justify}\par

\end{adjustwidth}

\begin{adjustwidth}{0.4in}{0.17in}
\begin{justify}

Avigad, Jeremy, and Brattka, Vasco. "Computability and analysis: the legacy of Alan Turing." (2014): 1-47 \end{justify}\par
\end{adjustwidth}

\vspace{\baselineskip}
\begin{adjustwidth}{0.4in}{0.17in}
\begin{justify}

Becher, Veronica, and Figueira, Santiago. "An example of a computable absolutely normal number." Theoretical Computer Science 270.1-2 (2002): 947-958\par
\end{justify}\par
\end{adjustwidth}

\vspace{\baselineskip}
\begin{adjustwidth}{0.4in}{0.17in}
\begin{justify}
Coron, Jean-Sebastien, and Naccache, David. "An accurate evaluation of Maurer's universal test." International Workshop on Selected Areas in Cryptography. Springer, Berlin, Heidelberg\  (1998) \par
\end{justify}\par
\end{adjustwidth}

\vspace{\baselineskip}
\begin{adjustwidth}{0.4in}{0.17in}
\begin{justify}
Di Gianantonio, Pietro. "A functional approach to computability on real numbers." Bulletin-European Association For Theoretical Computer Science 50 (1993): 518-518\par
\end{justify}\par
\end{adjustwidth}

\vspace{\baselineskip}
\begin{adjustwidth}{0.4in}{0.17in}
\begin{justify}
Gusﬁeld, Dan “G{\"o}del for Goldilocks: A Rigorous, Streamlined Proof of (a variant of) Gödel’s First Incompleteness Theorem” arXiv:1409.5944
\end{justify}\par
\end{adjustwidth}

\vspace{\baselineskip}
\begin{adjustwidth}{0.4in}{0.17in}
\begin{justify}
H{\aa}stad, Johan (2001). "Some optimal inapproximability results". Journal of the ACM. 48 (4): 798–859
\end{justify}\par
\end{adjustwidth}

\vspace{\baselineskip}
\begin{adjustwidth}{0.4in}{0.17in}
\begin{justify}
Khot, Subhash (2002). "On the power of unique 2-prover 1-round games". In Proc. 34th ACM Symposium on Theory of Computing
\end{justify}\par
\end{adjustwidth}

\vspace{\baselineskip}
\begin{adjustwidth}{0.4in}{0.17in}
\begin{justify}
Massey Jr, Frank J. "The Kolmogorov-Smirnov test for goodness of fit." Journal of the American statistical Association 46.253 (1951): 68-78\par
\end{justify}\par
\end{adjustwidth}

\vspace{\baselineskip}
\begin{adjustwidth}{0.4in}{0.17in}
\begin{justify}
Maurer, Ueli M. "A universal statistical test for random bit generators." Journal of cryptology 5.2 (1992): 89-105 \par
\end{justify}\par
\end{adjustwidth}

\vspace{\baselineskip}
\begin{adjustwidth}{0.4in}{0.17in}
\begin{justify}
Tao, Terrence. "Analysis I: 2nd ed." Hindustan Book Agency (2009): 185,195   
\end{justify}\par
\end{adjustwidth}

\vspace{\baselineskip}
\begin{adjustwidth}{0.4in}{0.17in}
\begin{justify}
Turing, Alan M. "On computable numbers, with an application to the Entscheidungsproblem." Proceedings of the London mathematical society 2.1 (1937): 230-265\par
\end{justify}\par
\end{adjustwidth}

\vspace{\baselineskip}
\begin{adjustwidth}{0.4in}{0.17in}
\begin{justify}
Wald, Abraham, and Wolfowitz, Jacob. "On a test whether two samples are from the same population." The Annals of Mathematical Statistics 11.2 (1940): 147-162.\par
\end{justify}\par
\end{adjustwidth}

\vspace{\baselineskip}
\begin{adjustwidth}{0.4in}{0.17in}
\begin{justify}
Weihrauch, Klaus. $``$A simple introduction to computable analysis$"$ Fernuniv., Fachbereich Informatik (1995)\par
\end{justify}\par
\end{adjustwidth}

\vspace{\baselineskip}

\vspace{\baselineskip}

\end{document}